\documentclass[runningheads]{llncs}

\usepackage{graphicx}
\usepackage{natbib}
\usepackage{array}
\usepackage{hyperref}
\usepackage{amsmath}
\usepackage{tikz}

\newcommand\apjs{ApJS}  
\newcommand\apj{ApJ}    
\newcommand\aap{A\&A}  
  
\newcommand\aaps{A\&AS}
\newcommand\aj{AJ}  
\newcommand\apss{Ap\&SS}  
\newcommand\apjl{ApJ Lett.} 
\newcommand\nat{Nature}   
\newcommand\mnras{MNRAS}  

\begin{document}
\title{Intraday variability of the polarization vector in AGN S5 0716+714}

\author{
        E.\,Shablovinskaya \inst{1} 
      \and 
        V.\,Afanasiev
       }

\institute{
           Special Astrophysical Observatory RAS, Nizhny Arkhyz, Russia, \email{e.shablie@yandex.com}
          }
\authorrunning{Shablovinskaya et al.}

\maketitle              
%



\begin{abstract}
The bright radio source S5 0716+714, which is usually classified as a BL Lac object,
is one of the most intensively studied blazar. S5 0716+714 demonstrates extremely
peculiar properties, such as the shortest time-scale of optical and polarimetric variations observed in blazars. In the given talk, we present the results of a 8-h polarimetric monitoring of S5 0716+714 with a $\sim70$-sec resolution carried out using the 6-m telescope BTA of the SAO RAS. The observation data analysis reveals the variability both in total and polarized light on the 1.5-hour timescales that specifies the size of the unresolved emitting region. The numerical model of polarization in jet with helical structure of magnetic field is suggested, and fitting the model reveals a magnetic field precession with a period of about 15 days.
\end{abstract}

%
\section{Introduction}

\label{intr}
BL Lac type objects or blazars\footnote{{Though the terms "BL Lacs" and "blazars" are not equal to each other, within this paper we would assume it interchangeably.}} are a special type of active galactic nucleus (AGN) with the jet directed almost toward the observer. Because of such orientation, the synchrotron (non-thermal) component formed in the jet makes a very large contribution in the blazar optical radiation, and short-term brightness and polarization variations are observed (the polarization degree in flashes is up 40\% and higher, see e.g. polarization light curve of PKS 1510-089 in the paper by \citet{PKS}).

S5 0716+714 is considered to be a typical BL Lac object. It shows flat power-law spectrum ($\alpha \geq -0.5, S_{\nu} \propto \nu_{\alpha} $ in the radio band), as well as intraday variability in all spectral ranges: from 8-12 hours in radio band \citep[6 cm,][]{Gorshkov2011a, Gorshkov11b} to 2.2-3.2 min flares in X-rays \citep{Pryal}. The observed polarization is also variable as within the night \citep[e.g.][]{Impey00, Amir06}, and on the scale of tens of days \citep[e.g.][]{Lar13}. However, the increase in the polarization is not correlated with optical flashes. 

It is assumed the plasma motion in a relativistic jet is responsible for the observed object variability. In this regard, this work is devoted to the detection of the polarization vector motion associated with the plasma, as well as the construction of a model of polarization changes within the night.

\section{S5 0716+714 –- what are you?}

The biggest question related to object S5 0716+714 is the redshift estimation. The lines weakness and their small equivalent width in the blazar spectra are normal since the contribution of the non-thermal jet component is too large. However, in the case of S5 0716+714, no details up to 0.3\% are detected in the spectrum that is also mentioned in the work by \citep{Nil08}.

\begin{figure}
    \centering
    \includegraphics[angle=90, scale=0.48]{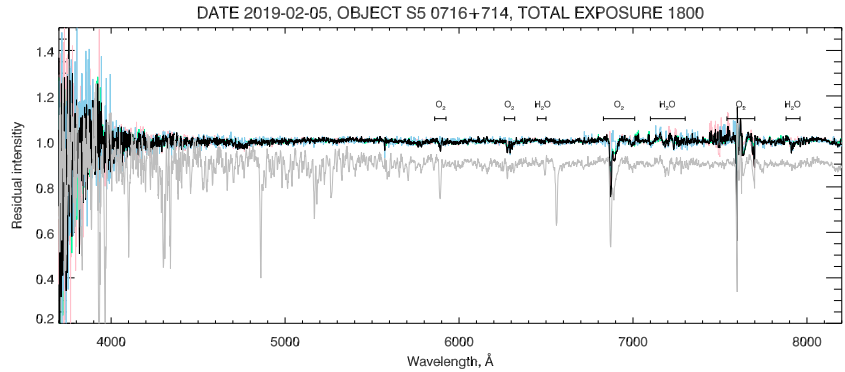}
    \caption{The spectrum of S5 0716+714 in residual intensities obtained while the object was in inactive state: the average S5 0716+714 spectrum is plotted in black and the comparison star is in gray. The telluric O$_2$ and H$_2$O lines are marked.}
    \label{spec}
\end{figure}

In February the spectroscopic observations of S5 0716+714 were carried out at the BTA telescope using the SCORPIO-2 device \citep{Af17}  while the object was in an almost record low activity state, when its luminosity was about 14.8 mag in $R$ band\footnote{According to the monitoring provided by Saint Petersburg State University observatory http://vo.astro.spbu.ru/en/program.}. Three 600 sec exposure spectra were obtained with a VPHG940@600 grating. The slit was oriented so that the spectrum of the neighbouring star was observed simultaneously with the object. In Fig. \ref{spec} the object and the star spectra are presented in the range 3700-8200\AA\ in residual intensities. As it can be seen from the figure, in the spectrum of S5 0716+714 there is no detail, despite low activity state, except noted atmospheric lines common to the object and the stars.

\begin{figure}
    \centering
    \includegraphics[scale=0.6]{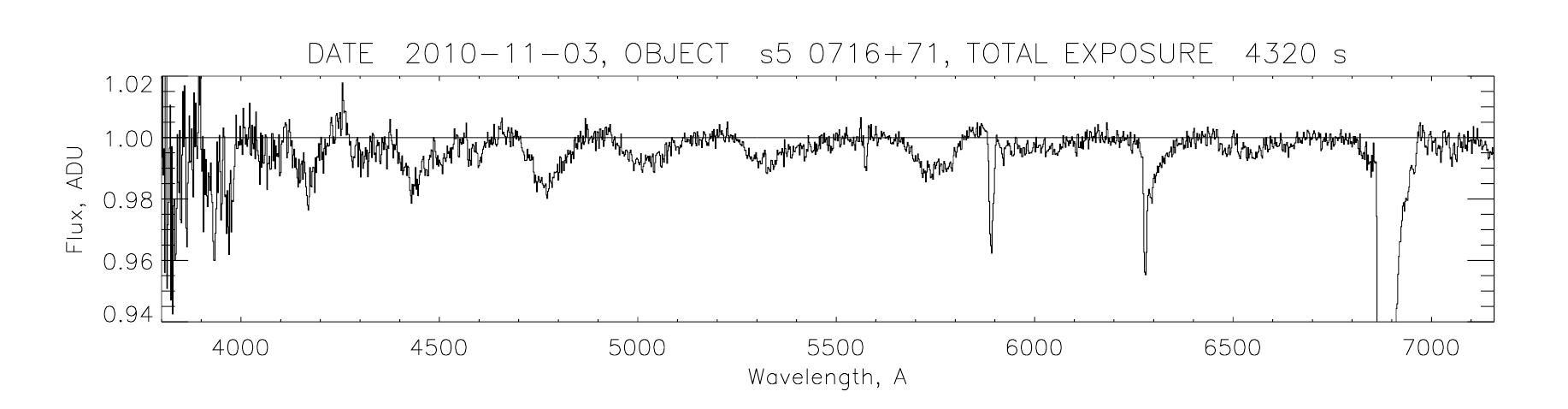}
    \includegraphics[scale=1]{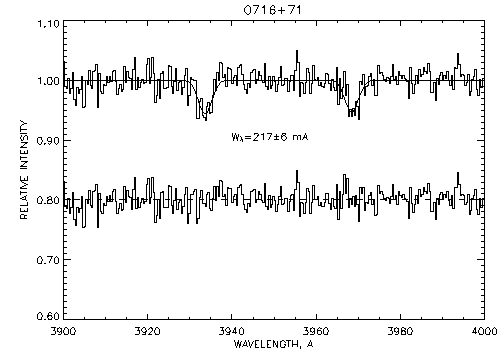}
    \caption{{\it Upper panel}: the spectrum of S5 0716+714 in residual intensities with DIBs. {\it Bottom panel}: equivalent width of Galactic H\&K CaII in S5 0716+714 spectrum.}
    \label{spec10}
\end{figure}

Also, the spectrum obtained at the BTA with the SCORPIO device in 2010 is shown (Fig. \ref{spec10}, upper panel). The object was in a brighter state ($\sim$13 mag in the $R$ band), but its lines are still not visible. The spectrum contains interstellar molecular bands (DIBs) and H\&K CaII lines  (Fig. \ref{spec10}, bottom panel). The H\&K CaII equivalent width is $W_{\lambda}=185 \pm 6$ m\AA, and according to the calibration \citep{Beers} is insufficient for an extragalactic object and corresponds to a distance less than 1 kpc. 

Other indirect methods were also used to find the S5 0716+714 redshift. The first try was described in the paper \citep{Stickel93}. Two weak galaxies (at distances of 27$''$ and 55$''$ from the source - 0.11 Mpc and 0.23 Mpc for $z \sim 0.3$, respectively) with redshifts of $\sim$0.26 were found and it was concluded the source $z$ is close to this value. However, the assumption about the galaxy cluster is not confirmed by X-ray data (see e.g. X-ray galaxy cluster surveys by \citet{Romer,Burenin}). Attempts were also made to detect the host galaxy. In the works \citep{Nil08, Stad14} the PSF fitting was used giving an inconsistent result\textbf{: $z \sim 0.3$ and }0.127 respectively. It is important to note that in the BL Lac type objects survey made by {\it HST} \citep{urry00} no evidence of the host galaxy was detected in S5 0716+714 despite the inactive state (14.18 mag in $R$), and a restriction on the redshift $z>0.5$ was given. 

Another specific feature of the S5 0716+714 object is its brightness. Based on the {\it HST} survey, the maximum brightness difference between the core and the host is 4 mag {on average} for the blazar sample. Yet for S5 0716+714 it is up 7 mag. The reason for such a tremendous brightness remains unclear.

The above-mentioned features of the object doubt on its extragalactic nature. {The observed synchrotron radiation is typical for all accreting systems both active nuclei and compact galactic systems, for example, low-mass X-ray binaries. In fact, when the system is oriented so that the relativistic jet points toward the observer and such details are absent as any marks of the surroundings or spectral features it is hard or even impossible to unambiguously answer the question about its type. This fact makes us free to raise an issue if S5 0716+714 is an extragactic object or it is just a system with a jet in the Galaxy, and we are looking for a critical test. }

Within the frames of this work, we will consider the object S5 0716+714 belongs to the BL Lac type, although the conclusions we give about the jet radiation can be applied to sources of a different nature.

\section{Polarimetric observations}

\begin{figure}
    \centering
    \includegraphics[width=0.35\textwidth,angle=90]{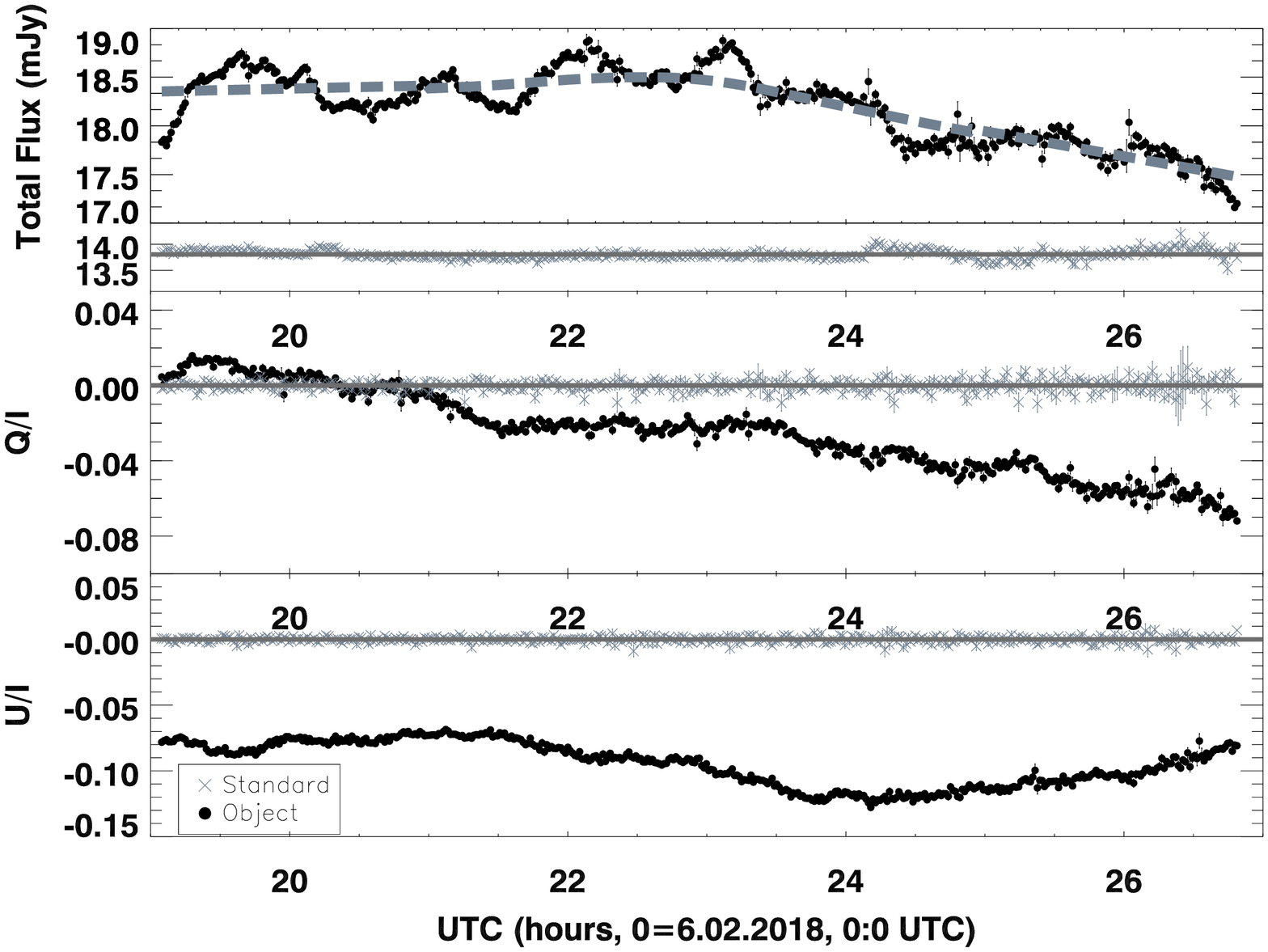}
    \includegraphics[width=0.35\textwidth,angle=90]{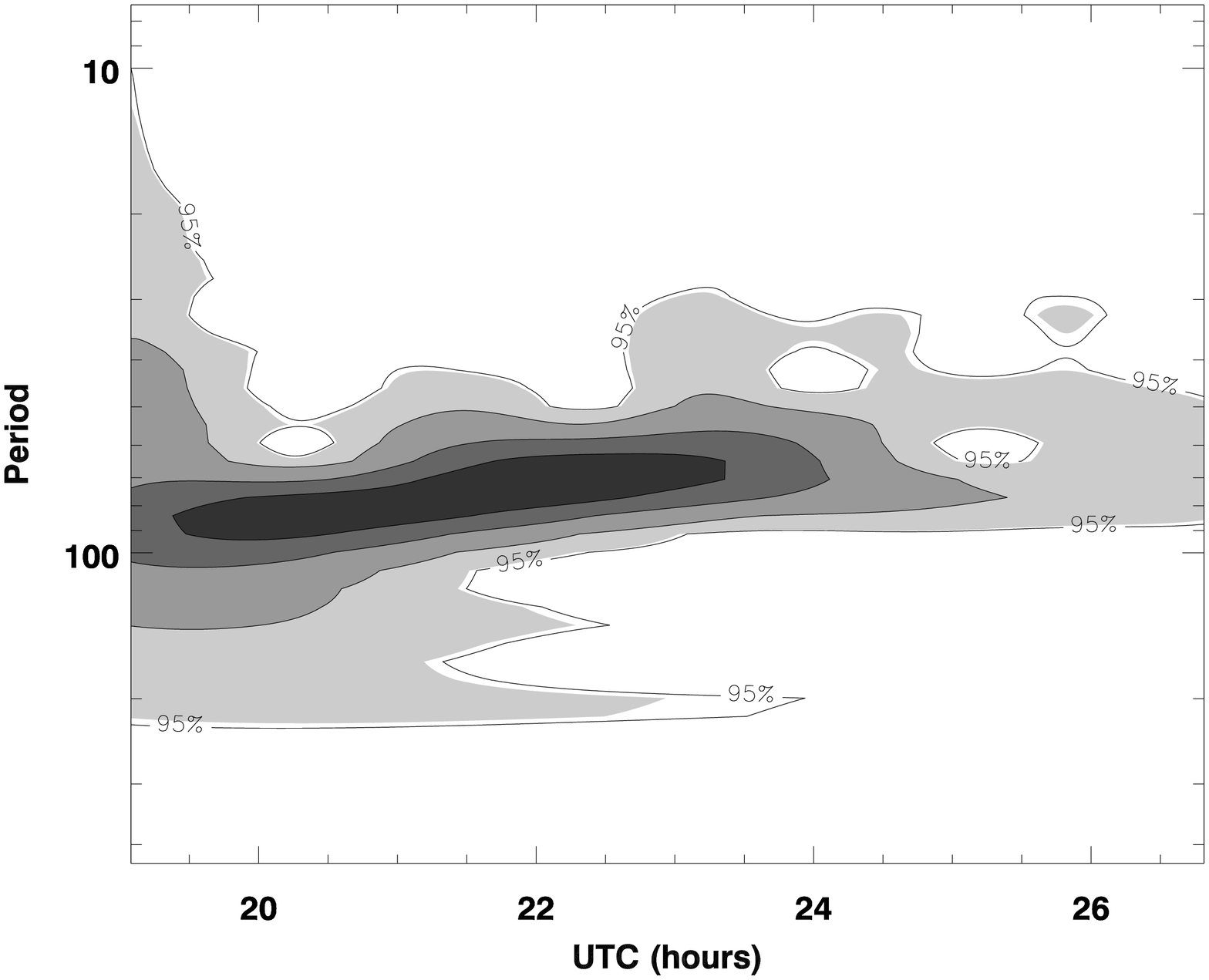}
    \caption{{\it Left}: the variations of the total flux and the Stokes parameters $Q$ and $U$ during the night. The observations started at 19:08 on 2018 February 2 (UTC). {\it Right}: The magnitude of the wavelet transform for the total flux.}
    \label{FQU}
\end{figure}

In February 2018, we conducted 8-hour polarimetric monitoring of the S5 0716+714 object with a 1-minute temporal resolution at the BTA telescope with the SCORPIO-2 device. There are two important methodic features:
\begin{itemize}
    \item[(i)] the observations of the object and the comparison star at a distance of $\sim$1$'$ occur simultaneously. Since the star is photometrically constant and has zero polarization \citep{Amir06}, it is possible to minimize the transmission variations and the atmospheric depolarization;
    \item[(ii)] the double Wollaston prism \citep{Geyer96, oliva97} was used as a polarization analyzer to measure both linear polarization parameters -- the Stokes parameters $Q$ and $U$ -- simultaneously.
\end{itemize}
The obtained accuracy was about 0.005 mag for photometry and 0.1\% for polarimetry.

As a result, an 8-hour data series was obtained (Fig. \ref{FQU}, left), where significant changes in both flux and polarization are observed. To study the total flux variability the long-period trend was approximated {by the robust smoothing 2-degree polynomial function\footnote{The function was close to 6-degree polynomial one.} }and subtracted. {The wavelet analysis \citep{wavelet}} provided a period of $77\pm10$ min of rapid variations (Fig. \ref{FQU}, right). 

\begin{figure}
    \centering
    \includegraphics[scale=0.4,angle=90]{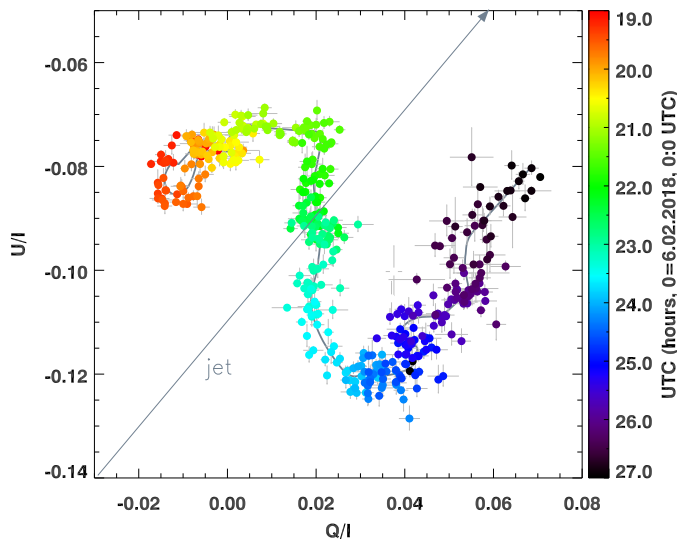}
    \caption{The variations of the normalized Stokes parameters $Q$ and $U$
during the night on the $QU$-diagram.}
    \label{dra}
\end{figure}

To study the variability of the $Q$ and $U$ parameters, they were plotted on the $QU$-plane (Fig. \ref{dra}). During the night the polarization vector changed its direction several times about every 1.5-3 hours, and the changes are perpendicular to the jet direction. Therefore, "loops" and "arcs" are observed on the $QU$-plane. Moreover, the period of the polarization vector direction and the total flux changes are similar. Indeed, if we assume that the observed motion of the polarization vector is caused by the plasma motion in the jet, then the polarization vector rotation will be due to the plasma changing direction and, consequently, a change in the Doppler amplification of its brightness for the observer.

\section{Data comparison and model}

Similar results were found for S5 0716+714 in earlier papers. Thus, the discovery that the polarization of an object changes on the $QU$-plane not stochastically but along the definite trajectories was made in the work \textbf{by} \citet{Impey00}. However, the time sets duration was not enough to investigate the way the polarization vector change. 

Furthermore, in the case of the object BL Lacertae (ancestor of the class blazars), a similar picture was found. We have examined polarimetric data, obtained in the paper{ by} \citet{covino15} as the angle and degree of polarization separately, on the $QU$-plane, where we also obtained the rotation of the polarization vector on the scales of several hours. Besides, for BL Lac, such a conclusion was made earlier in the almost forgotten work \cite{moore82}: BL Lac showed the rotation of the polarization vector on the scales of hours in more than 7-day monitoring. 

Also, the rotation of the polarization vector along the arcs was found in the radio band in observations of blazar CTA 102 \citep{li18}. It was assumed there that the plasma in the jet should rotate along helical trajectories. On the other hand, according to the commonly used model \citep{mar08}, it is known that the optical synchrotron radiation in the AGN jet is formed at a distance of $\sim$10$^{-3}$ pc from the central source, where the magnetic field has a helical structure.

\begin{figure}
    \centering
    \includegraphics[scale=0.4,angle=90]{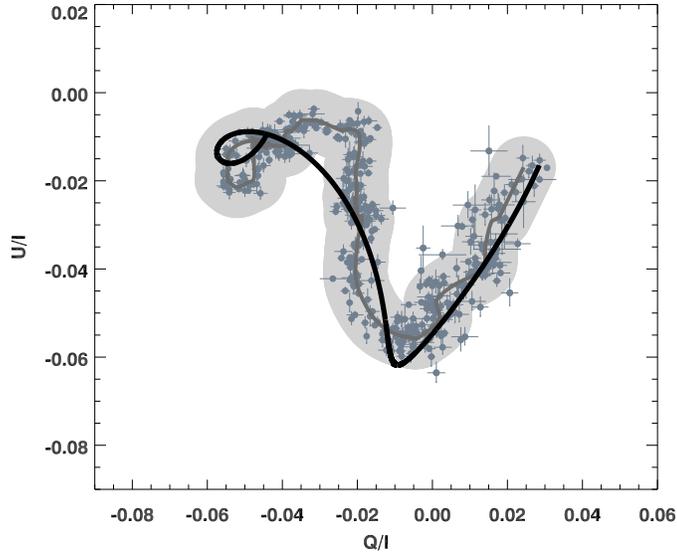}
    \caption{The model of linear polarization in the $QU$-plane in the case of
jet precession. The observational data are plotted with grey dots with error
bars. The 3$\sigma$ confidence area is coloured light grey.}
    \label{mod}
\end{figure}

Based on these principles, we constructed a geometric model of polarization change in the jet helical magnetic field \citep[in more details -][]{SA19}. However, the numerical model of polarization change showed that it is impossible to describe the observed motion of the polarization vector by a stable configuration of the field. An important feature of our model was the field precession as an additional kinematic component. Then the rotation of the polarization vector on the $QU$-plane observed during the 8-hour monitoring is described by the motion of the plasma in a helical magnetic field precessing with a period of $\sim$15 days. Moreover, the linear size of the region where the optical polarization is formed is associated with the time of its variability, that is, they are about 1.5 lt hours or 10 a.u. Comparison of the model with observational data is shown in Fig. \ref{mod}, {where $3\sigma$ confidence area of the smoothed polarization vector rotation is plotted light grey. The model fits the data good except the region between 21 and 22 hours. We attribute this divergence to the
physical processes (radiation transfer and etc.) that we have not considered as our model describes the geometry of the plasma motion, and this neglect does not affect our quantitative results.}

\section{Conclusions}
During the polarimetric monitoring of the S5 0716+714 object  we obtained the following results.
\begin{itemize}
    \item[(i)] We found the variability of the total ($\Delta$=0.04 mag) and polarized ($\Delta$=7\%) fluxes on a time-scale $\sim$1.5 hours.
    \item[(ii)] We discovered the specific pattern of the polarization vector on the $QU$-plane -- "arches" and "loops".
    \item[(iii)] The estimation of the linear size of the field identifying with the emitting region – $1.5 \cdot 10^{-5}$ pc, or 10 a.u. at $\sim$10$^{-3}$ pc from the central BH.
    \item[(iv)] The polarization vector rotations mark the magnetic field precessing with the 15 days period.
    \item[(v)] The similar pattern was found in other papers and also for BL Lac.
\end{itemize}


\section*{Acknowledgements}
We sincerely thank V. R. Amirkhanyan for valuable discussions
and useful remarks. The results of observations were obtained with
the 6-m BTA telescope of the Special Astrophysical Observatory
Academy of Sciences, operating with financial support from the
Ministry of Education and Science of Russian Federation.

\bibliographystyle{mnras}
\bibliography{lit}

%

\end{document}